\documentclass[prl,preprint]{revtex4}

\usepackage{graphicx}

\begin{document}

\title{Spectral diffusion of single semiconductor nanocrystals: the
influence of the dielectric environment}

\author{Daniel E. G\'omez, Joel van Embden, 
        and Paul Mulvaney\footnote{Corresponding author, email:
            mulvaney@unimelb.edu.au}} 

\affiliation{School of Chemistry, 
             The University of Melbourne, 
             Parkville, VIC 3010, 
             Australia} 

\date{\today}

\begin{abstract}
                                
  We have explored the influence of different matrices on the emission
  line shape of individual homogeneously coated CdSe/CdS/ZnS
  nanocrystals. The results obtained corroborate previous observations
  of a correlation between blinking events and spectral diffusion but
  in addition we have found that the extent of spectral diffusion is
  almost independent of the dielectric environment of the
  NC. Additionally, we report the observation of a correlation between
  the line width and emission energy which is not expected to occur in
  the spherical - symmetric NCs employed in this work. The
  implications of these results are discussed.

\end{abstract}

\maketitle

Chemically synthesized semiconductor nanocrystals (NCs) exhibit
interesting optical properties when studied individually. For
instance, the photoluminescence (PL) emission experiences continuous
random shifts both in its peak position and line width, a process
known as spectral diffusion \cite{empedocles-jpc}. Stark spectroscopy
experiments carried on individual NCs \cite{empedocles-science} have
revealed that the emission peak position can shift as the result of
the application of an external electric field that modifies the NC
energy-level spacing through the quantum-confined Stark effect (QCSE)
\cite{qcse}. Consequently, the observed spontaneous spectral diffusion
is thought to occur due to fluctuating electric fields produced by
accumulation and redistribution of charges placed either in the
near-by environment of the NC or directly on its surface
\cite{empedocles-science}. These fluctuations are believed to dominate
the line width of the emission of single NCs due to an inherent
averaging on the position of emission that takes place during the time
required to measure a spectrum \cite{empedocles-jpc}, explaining why
the widths obtained are significantly larger than the reported
homogeneous line width measured by spectral hole burning
\cite{palinginis}. In addition to these observations, recent
experiments have shown that there is a correlation between the extent
of these spectral shifts and switch - on events in the
photoluminescence blinking of single NCs \cite{neuhauser}. Blinking in
NCs has been explained in terms of an Auger recombination mechanism,
whereby a charged NC is in a non - emissive state due to efficient
energy transfer from exciton annihilation to the excess charge carrier
present in the NC \cite{nirmal}. Evidence for this model is supported
by the intermittent presence of charges on single NCs \cite{krauss}
and the sensitivity of blinking dynamics to the dielectric environment
of the nanocrystal \cite{cichos}. This increases the probability of
observing longer off-times in matrices with high static dielectric
constants as a result of better solvation of the charged NC.

Based on this experimental evidence and the observed correlation
between blinking and spectral diffusion, it seems reasonable to
question whether there is a relationship between the dielectric
environment of the NCs and spectral diffusion. In this paper we
address this issue by following the temporal evolution of the emission
spectra of single NCs embedded in matrices with different dielectric
constants. We show that the extent of spectral diffusion is not
affected by the polarity of the external medium that surrounds the
NCs, and we show that for {\it spherically symmetric} NCs, the position
of the PL is correlated to its line width.

\begin{figure}
\includegraphics{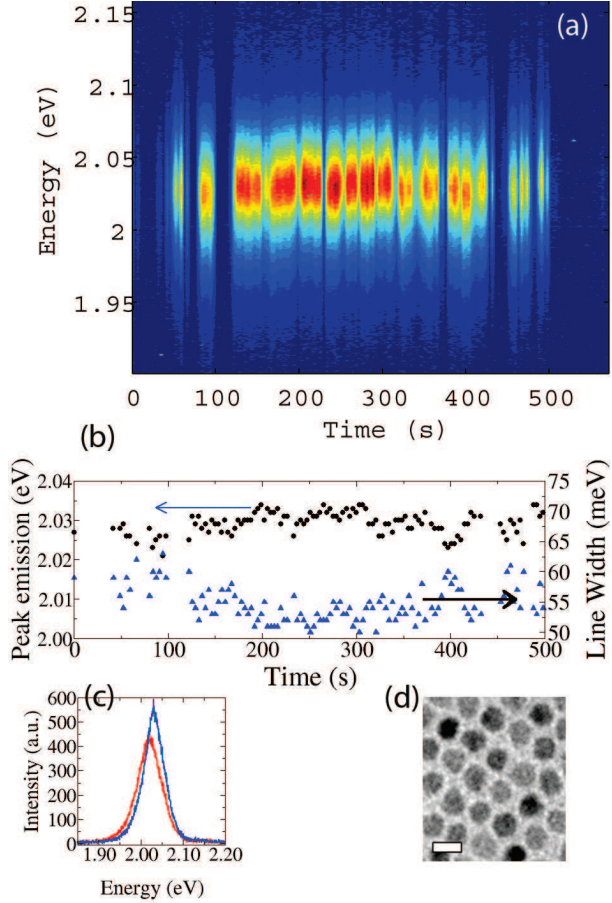}
\caption{Spectral diffusion of the emission of a single CdSe/CdS/ZnS
  (4.3 nm core diameter, ensemble emission max. 2.019 eV) NC measured
  at room temperature in PMMA. (a) Temporal evolution of the PL
  spectrum. (b) Peak position and line widths as obtained from
  Lorentzian fits to the data shown in (a). (c) Switch - on spectral
  shifts (d) TEM image of the sample used, scale = 10 nm. Excitation
  power = 300 nW, integration time 3s. Excitation wavelength 488nm}
\label{fig1}
\end{figure}

We studied the photoluminescence (PL) of single CdSe/CdS/ZnS
core/shell NCs that were synthesized following the methods of van
Embden {\it et al} \cite{joel, joel2} (core diameter 4.3 nm, overall
shell thickness 2 nm, passivated with octadecylamine and tri-butyl
phosphite, ensemble emision max. 2.019 eV). A representative TEM image
shown in Fig \ref{fig1}(d). The samples were highly diluted from
chloroform dispersions and spin-coated on top of thin films of polymer
matrices that were initially deposited on clean glass coverslips from
1 \% w/w solutions. For the preparation of these films, we employed
poly(vinyl alcohol) (PVA), poly(N-vinylpyrrolidone) (PVP), poly
(methylmethacrylate) (PMMA) and polystyrene (PS). PMMA and PS were
dissolved in chloroform together with the NCs, PVP was dissolved in
ethanol whereas PVA was dissolved in demineralized water. The PL of
individual NCs was measured using a modified commercially available
laser-scanning confocal microscope (Olympus Fluoview 500), using a
60x/1.4 numerical aperture objective in the epi - illumination
configuration. The PL was excited with the 488 nm line of an argon ion
laser at typical powers of 300 nW (excitation power measured in front
of the microscope objective) and dispersed onto a 0.55 m imaging
spectrograph with a resolution of 0.1 nm (Jobin Yvon, TRIAX 550).  The
spectra were acquired at room temperature with integration times of 3s
using a liquid-nitrogen-cooled charge-coupled device camera.

In Figure \ref{fig1}(a) we show a series of consecutive PL spectra for
a single NC in PMMA. Consistent with previous reports
\cite{empedocles-jpc, muller-prl, muller-prb}, we observe that both
the peak position and the line width of the emission of an individual
NC constantly change during the measurement time
(Fig. \ref{fig1}(b)). The extent of these changes can have a net value
of up to 10 meV for the peak position, with the direction of these
shifts fluctuating randomly in time. This demonstrates that the
spectral shifts are not due to an irreversible process, such as
photoinduced chemical changes or permanent reorganizations at the
surface of the NCs. Additionally, it can be seen in this figure that
relatively big shifts in the PL peak position are always likely to
follow a switch-on event, a result that confirms the existence of a
correlation between blinking and spectral diffusion \cite{neuhauser}.

\begin{figure}
\includegraphics{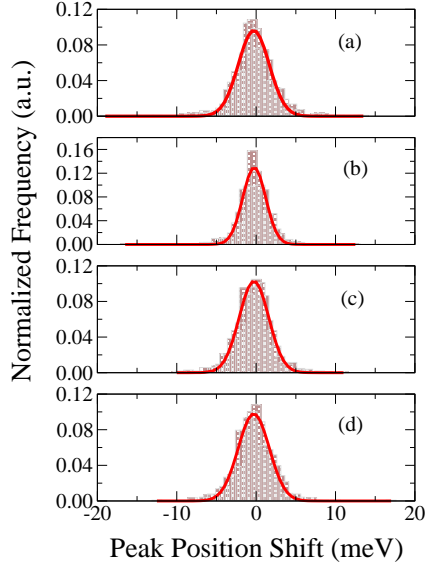}
\caption{Histograms of the difference in emission energy
(E$^{max}_{i+1}$ - E$^{max}_i$) between consecutive emissive events
compiled from the emission of single CdSe/CdS/ZnS NCs deposited in (a)
PS, (b) PMMA (c) PVP (d) PVA. The lines are best fits to a Gaussian
distribution.}
\label{fig2}
\end{figure}

In order to assess the possible effects of the matrix on the observed
SD, we compiled histograms of the energy differences between
consecutive emission events for a set of NCs deposited in different
polymer matrices; the results are shown in Figure \ref{fig2}. In
agreement with the works in references \cite{neuhauser, muller-prl,
muller-prb}, the histograms of Figure \ref{fig2} can be described by
Gaussians, the width of which serves as an indication of the extent of
spectral diffusion.  As can be seen in Table \ref{table}, the values
obtained for the widths of these histograms do not differ
significantly for the different matrices employed. This indicates that
the extent of spectral diffusion during consecutive emission events
cannot be linked to charge trapping and reorganization in the matrix,
as such processes would be sensitive to dielectric stabilization
\cite{cichos}.  As discussed in reference \cite{neuhauser}, there are
shifts in emission energy that occur in conjunction with blinking
events which are known to take place on broadly distributed time
scales \cite{kuno}. Given that the histograms of Figure \ref{fig2}
were compiled from sets of spectra acquired on a time scale of 3
seconds, the widths of these histograms also include time-averaged
information of events in which the NC, according to the charging
hypothesis, could have ejected a charge carrier into the matrix
several times. On this basis, the results shown in Table \ref{table}
demonstrate that ionization of the NC is not the dominant mechanism
responsible for the magnitude of the SD and that if charge
reorganization is responsible for the spectral shifts,then it must
take place either at the surface of the NCs or directly at the
core-shell interface. We note additionally that for the case of PMMA,
the obtained width of 3.7 meV is comparable to that reported by
Neuhauser {\it et al} from similar experiments that were carried out
at cryogenic temperatures \cite{neuhauser}. The fact that the width of
this diffusion histogram remains almost unchanged by an order of
magnitude increase in temperature indicates that the mechanism
responsible for the extent of spectral diffusion is not thermally
activated \cite{muller-prl, muller-prb}.

\begin{figure}
\includegraphics{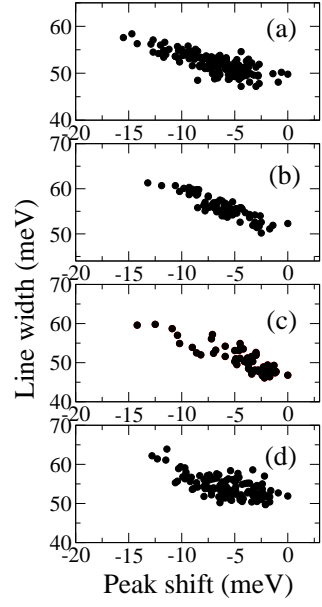}
\caption{Correlation between line width and peak position shift for the
emission of single CdSe/CdS/ZnS NCs in (a) PS, (b) PMMA, (c) PVP and (d) PVA.}
\label{fig3}
\end{figure}

In Figure \ref{fig3} the emission line width and peak position shift
(defined as the difference between the peak emission energy and the
highest emission energy recorded in a given time series) are plotted
for single NCs deposited in the different matrices. This figure
clearly shows a correlation between these two parameters for spherical
core-shell NC: the emission line width increases as the spectra
redshifts.

This type of correlation has recently been observed in CdSe NCs
overcoated with an elongated CdS shell \cite{muller-prl,
muller-prb}. For this system, it was postulated that a QCSE is
produced by a surface charge that moves and oscillates along the
shell. Due to the one - dimensional character of this shell, the
strength of the electric field at the NC core depends on the location
of this surface charge along the length of the shell. When the charge
is located close to the core the electric field perturbing the exciton
increases, the effect of which is twofold. Firstly, the emission peak
position is redshifted and secondly a broadening of the emission
profile is observed due to fluctuations in the peak position, which
occur to a greater extent in the presence of higher field strengths
\cite{muller-prl, muller-prb}.  However a corollary of this hypothesis
is that such correlations {\it should not be observed} for spherically
symmetric NCs, since in this case the distance from the NC core to the
excess charge is radially symmetric. The results presented in Figure
\ref{fig3} clearly demonstrate that fast oscillations and diffusion of
a charge carrier along the length of an elongated shell are not
responsible for the broadening associated with the redshifted emission
and that such correlations are intrinsic to the NCs, regardless of the
symmetry of their capping shells. This have been supported not only by
the data presented here but also by the fact that similar observations
have been reported for CdS nanorods in the absence of an inorganic
shell \cite{kulik}.

In summary, we have presented experimental data that shows that the
extent of SD does not depend on the dielectric constant of the matrix
surrounding the NCs. It is therefore possible to infer that the charge
reorganization that leads to the QCSE responsible for the spectral
shifts occurs either at the surface of the NCs or directly at the
core-shell interface. We have also shown the existence of a
correlation between the position of the emission and the line
broadening for NCs of spherical shape. This suggests that the
mechanism responsible for the correlated broadening and redshift in
the PL is intrinsic to the NC core. However,the exact mechanism that
gives rise to the correlation remains at present unknown.

This work was supported by
ARC Discovery Grant DP0451651.

\begin{table}
\begin{center}
\caption{Widths of Gaussian fits to the histograms shown in figure
\ref{fig2}. The values in parentheses correspond to the uncertainties
of the fit. The static dielectric constants, $\epsilon_0$, were
obtained from ref. \cite{cichos}.}

\begin{tabular}{l c c}
\hline
\hline
Matrix    & {$\epsilon_0$} & {width (meV)} \\ \hline
{PS}      & 2.5            &  3.9 (0.1)    \\ 
{PMMA}    & 3.4            &  3.7 (0.1)    \\ 
{PVP}     & 4.8            &  2.9 (0.2)    \\ 
{PVA}     & 14             &  3.9 (0.1)    \\ \hline
\hline
\label{table}
\end{tabular}
\end{center}
\end{table}

\end{document}